\documentclass[11pt]{article}
\usepackage[dvips]{graphicx,graphics}
\usepackage{multicol,latexsym}
\usepackage{amsmath,amssymb}
\usepackage[usenames]{color}
\newcommand{\spc}{\hspace{0.22in}}
\newcommand{\spn}{\hspace*{0.22in}}

\newif\ifequationnumber
\def\eqswitchon{\equationnumbertrue
\@addtoreset{equation}{section}
\def\theequation{\arabic{section}.\arabic{equation}}
}
\def \ds {\displaystyle}
\def \fn {\footnotesize}
\def \ns {\normalsize}
\def \sp {\scriptsize}
\def \es {\enspace}
\def \ts {\thinspace}
\def \nts {\negthinspace}
\def \nms {\hspace*{-0.25cm}}

\title{
  \vspace*{-1.0cm}
  \hfill{\ns KEK-TH-958} \\ 
  \vspace*{-0.5cm}
  \hfill{\ns May 2004}  \\
  \vspace*{2.0cm}
  {\LARGE {\bf M-Theory Phenomenology}} \\
  \vspace*{0.3cm}
  {\LARGE {\bf and See-Saw Mechanisms}\footnote{Talk given at 
  {\sl Neutrino Mass and See-Saw Mechanism,\/} Fujihara Seminar
  held at KEK, Japan, Feb.~23-25 2004.}} \\
  \vspace*{1.2cm}
}
\author{{\sc Hirotaka Sugawara}\thanks{E-mail:{\tt sugawara@phys.hawaii.edu}.
Address and e-mail after April~1:~{\it Graduate University for Advanced 
Studies, \quad\quad International village, Hayama, Kanagawa-ken, Japan};\/
{\tt sugawara\_hirotaka@soken.ac.jp}.}}
\date{}
\begin{document}
\maketitle
\vspace{-1.0cm} 
\begin{center}
{\it Department of Physics and Astronomy, University of Hawaii at Manoa, \\
2505 Correa Road, Honolulu, Hawaii 96822}
\end{center}
\vspace{1.9cm}
%
%
\begin{abstract}
  A version of M-theory phenomenology is proposed in which the symmetry is 
based on the group $SO(10) \times SO(10) \times SO(10) \times U(1) \times 
U(1)$. Each $SO(10)$ group acts on a single generation. The $U(1) \times
U(1)$ is regarded as the hidden sector symmetry group. The supersymmetry is
broken in the hidden sector by the Fayet-Iliopoulos $D$-term for each group. 
The $D$-term is needed also to circumvent the powerful non-renormalization 
theorem since the $SO(10) \times SO(10) \times SO(10)$ is broken down to the 
usual $SO(10)$ by the pair condensation of certain messenger sector multiplets.
The exchange of $U(1)$ gauge bosons gives an attractive force for the pair to 
be created and condensed. The off-diagonal mass matrix elements among the 
generations in these messenger sector multiplets are the source of the flavor
dynamics including the CP violation. The pair condensation of another multiplet
in the messenger sector leads to the doublet-triplet splitting. The $SO(10)$ 
decuplet Higgs couples only to one of the generations. The other couplings 
should, therefore, be calculated as higher order corrections. We present 
our preliminary results on the calculation of the mass matrices and the mixing
angles for leptons and quarks in this model.
\end{abstract}
%
%
\newpage
%
\section{Introduction}
  There are two significant facts which imply the necessity of going to 
ultra-high energy.
\begin{itemize}
\setlength{\itemsep}{-2pt}
\item[(1)] Renormalization group \\
\spn As is shown in Fig.~\ref{fig:runCouplings}, the 
unification scale is as large as $10^{16}~{\rm GeV}$~\cite{renGroup}.
%
%
\begin{figure}[h]
\vspace*{-0.3cm}
\begin{center}
\setlength{\unitlength}{1mm}
\begin{picture}(75,73)(0,0)
\put(0.0,0.0){\includegraphics[width=7.5cm]{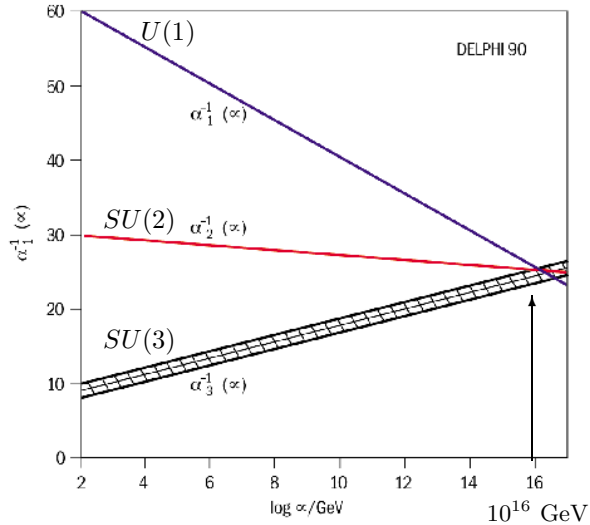}}
\put(18,64){\color{black}\small $U(1)$}
\put(13,39){\color{black}\small $SU(2)$}
\put(13,23){\color{black}\small $SU(3)$}
\put(70,8){\color{black}\vector(0,22){22}}
\put(64,00){\color{black}\fn $10^{16}~{\rm GeV}$}
\end{picture}
\end{center}
\vspace*{-0.8cm}
\caption{Running gauge coupling constants.}
\label{fig:runCouplings}
\end{figure}
\item[(2)] See-saw mechanism \\
\spn This is the theme of this workshop and it was suggested 30 years ago by 
Gell-Mann, Ramond and Slansky~\cite{GMRS} and by Yanagida~\cite{Yanagida} to 
explain the smallness of the 
neutrino mass. If we take the standard model scale to be $m=100~{\rm GeV}$ and
the neutrino mass to be $1~{\rm meV}$, the Majorana mass is as large as 
$10^{16}~{\rm GeV}$:
\begin{eqnarray}
\frac{\ds m^2}{\ds M} \sim \frac{\ds (100~{\rm GeV})^2}{\ds 10^{16}~{\rm GeV}}
= 10^{-12}~{\rm GeV} = 1~{\rm meV}\ts.
\label{eq:seasawMass}
\end{eqnarray}
These do not necessarily imply the Planck mass, but they are close enough. 
We might say that the string theory is vaguely implied by these facts since we
know that the string theory seems to be the only candidate at this stage to 
unify the gravity with other interactions. 
\end{itemize}
\vspace{0.3cm}

   I note that the older generation of physicists (for example, P.~Ramond) 
worked both in string theory and in phenomenology. But the new generation 
(apart from a few exceptions) works either in string or in phenomenology.
I would say that this is a very unhealthy condition! Here I would like to start
from the M-theory and try to go all the way down to calculate the quark-lepton 
mass matrices, mixing angles and so on. The present work is partly motivated
by a paper of E.~Witten~\cite{wittenG2}.

  The talk is organized in the following way. I will first very briefly touch 
on the subject of M-theory. Although the current conventional wisdom is that
there is no consistent way to formulate it as the quantum membrane theory, I 
would like to challenge this. The mathematical techniques I will rely on are
some classification methods of topology of 3-manifolds, such as the Heegaard
diagram~\cite{Heegaard} on one hand, and the remarkable conjecture by 
Thurston~\cite{Thurston} which tells us that the geometry of a 3-manifold is 
unique, given its topology, on the other. These powerful techniques at least 
encourage us to investigate the possible existence of the renormalizable and
unitary quantum membrane theory.

  Next, I will switch to the phenomenology by fixing the vacuum state of the
11-dimensional supergravity theory, taking into account the following 
facts:
\begin{enumerate}
\setlength{\itemsep}{-2pt}
\item There are three generations of quarks and leptons.
\item The $SO(10)$ symmetry is the correct gauge symmetry.
\end{enumerate}
We combine these facts with the theoretical consequences of M-theory:
\begin{enumerate}
\setlength{\itemsep}{-2pt}
\item Compactified space is the 7-manifold of $G_2$ holonomy~\cite{G_2holonomy}
which presumably has the structure of $K_3 \times \Omega$ where $\Omega$ is a 
certain 3-manifold and $K_3$ is a 4-manifold with $SU(3)$ holonomy.
\item M-theory gives a natural explanation of the 
``deconstruction''~\cite{deconstruction} which
means that the symmetry group of $K_3$ can be reached only from isolated
points in $\Omega$.
\end{enumerate}

  An explicit model with the symmetry of $\{SO(10)\}^3 \times 
\{U(1)\}^2$~\cite{explicitModel} is constructed in section~\ref{sect:expModel}.
Each $SO(10)$ corresponds to each of the three generations of quarks and 
leptons. $\{U(1)\}^2$ is thought to be the gauge symmetry of the hidden sector.
We explicitly specify the particle content of the hidden sector and the 
``messenger sector'' which is defined to be a set of particles which transform
non-trivially, both in the physical and the hidden sector groups. One of the 
most important ingredients
here is the fact that the $SO(10)$ decuplet Higgs couples only to one of the
generations. The rest of the couplings should be calculated as higher 
order corrections. This enables us to calculate the quark-lepton mass
matrices and the mixing matrices, starting from the first principles.

  Section~\ref{sect:symbrkHrchy} is devoted to the discussions on the broken
symmetries and the hierarchy issues. \\ The $\{U(1)\}^2$ for the hidden sector
is introduced for the following purposes:
\begin{enumerate}
\setlength{\itemsep}{-2pt}
\item Break the supersymmetry in the hidden sector using the Fayet-Iliopoulos
mechanism~\cite{FayetIliopoulos}.
\item Break the $\{SO(10)\}^3$ down to $SO(10)$ by dynamically breaking the
symmetry with the condensation of the Cooper pairs of messenger sector
multiplets.
\end{enumerate}
We circumvent the powerful non-renormalization 
theorem~\cite{nonRenrmlztnThrm} by using the Fayet-Iliopoulos $D$-term. One of
the hierarchies is provided by the masses and couplings in the original 
Lagrangian, presumably calculated using the techniques of the membrane 
instanton~\cite{BecBecSt}. Other kinds of hierarchical mechanisms are also 
important in our model. The see-saw mechanism is one example. Another mechanism
is given by the hierarchical nature of the BCS-NJL order 
parameter~\cite{hrchyBCS_NJL}. In fact, the doublet-triplet splitting is the 
consequence of this mechanism in our model combined with the possible membrane
instanton calculations.

  Section~\ref{sect:QrkLptnMtrx} describes the calculation of the mass matrices
and the mixing angles of quarks and leptons which I am carrying out with 
S.~Pakvasa~\cite{Pakvasa}. Although this work is still in progress, we can
already see some of the interesting features of the model. The Wolfenstein
parameterization for the quark mixings and the bi-maximal nature of the 
neutrino mixings can naturally be seen in our model. Part of CP violation can 
come from the non-zero phase of the BCS-NJL order parameters. 
The detailed calculation is underway.
%
\section{M-theory}\label{sect:M_theory}
\spc M-theory is a theory which is supposed to reduce to the 11-dimensional
supergravity theory in the low energy regime. The matrix formulation 
exists~\cite{mtrxFrmltn} which seems to be a renormalizable theory as long as 
the compactified space is appropriate~\cite{cptSpcmtrx}. We know this theory is
equivalent to the membrane theory in the classical case~\cite{mbrnThry}. 
Can the membrane theory be quantized as a renormalizable theory\/? The 
conventional wisdom seems to be negative. The membrane theory, then, is 
regarded as an effective theory, just as the other $n$-dimensional theories
with $n \ge 3$.
I am not quite satisfied with this situation, since the 3-manifolds which come
into play in the membrane theory have a lot in common with the 2-manifolds.
The most notable examples are:
\begin{itemize}
\setlength{\itemsep}{-2pt}
\item[(1)] The topology of the closed 3-manifolds can be 
classified~\cite{3-mfdTplgy} (but not uniquely) by a method which seems to be
an extension of the genus expansion in the 2-manifold case, i.e.~the Heegaard 
expansion.
\item[(2)] There seems to be a unique geometry up to homeomorphism for a 
3-manifold of a given topology, i.e.~Thurston conjecture~\cite{Thurston}.
\end{itemize}
  These two facts (one is still a conjecture) are at least very encouraging
for tackling the problem of renormalizable membrane
theory.

  Now, the closed membrane vacuum amplitude is given by
\begin{eqnarray}
\langle 0 | S | 0 \rangle = \sum_{\mbox{\sp all closed} 
\atop \mbox{\sp 3-manifolds}} \int {\cal D} g_{ij} {\cal D} X_i
\exp\left(- \int {\cal L} d^3 \sigma \right) \es,
\label{eq:clsdMembrnVacAmp}
\end{eqnarray}
with $g_{ij}$ the metric of a 3-manifold and $X_i$ the coordinate for the
manifold in the 11-dimensional space-time. The sum over all closed manifolds 
means over both topology and geometry. The topological aspect has been 
investigated in my earlier paper~\cite{sugawaraMnbrn}. We introduce the
idea of topology and geometry of 3-manifolds in what follows.
%
\subsection{Topology}
\spn Heegaard diagram:~any 3-manifold can be expressed in one or more than
one Heegaard diagrams~\cite{Heegaard}. 
%
%
\begin{figure}[h]
\vspace*{-0.7cm}
\begin{center}
\setlength{\unitlength}{1cm}
\begin{picture}(12,6.0)(0,0)
\put(0.0,0.0){\includegraphics[width=12cm]{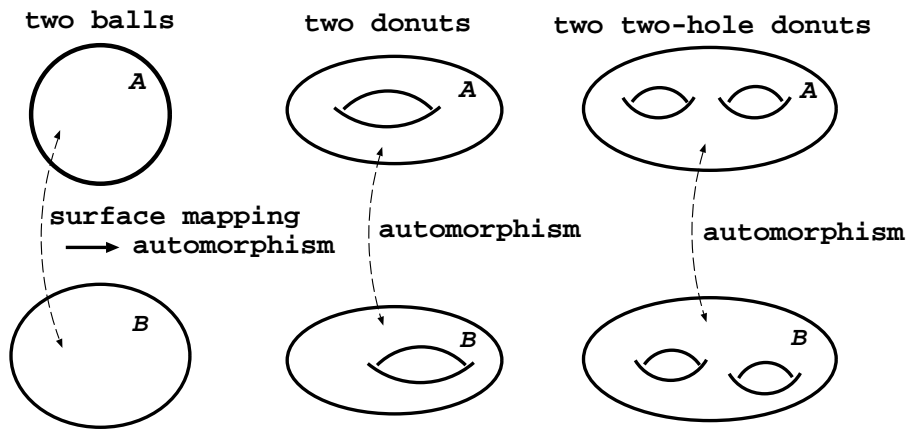}}
\end{picture}
\end{center}
\vspace*{-0.8cm}
\caption{Heegaard diagrams}
\label{fig:srfcAtmrphsm}
\vspace*{-1.0cm}
\end{figure}
\clearpage
\noindent{Fig.~\ref{fig:srfcAtmrphsm}} shows some examples of the Heegaard
diagrams. Here the classification of the surface automorphism gives rise to the
classification of topology of the 3-manifolds. The following are the two
simple cases:
\begin{enumerate}
\setlength{\itemsep}{0pt}
\item \underline{Two-balls}
\vspace*{-2mm}
\begin{eqnarray*}
S^3 
\nts&:&\nts x^2 + y^2 + z^2 = 1\\
{\rm (A)}
\nts&:&\nts x^2 + y^2 = 1 - z^2 \le 1,\es z \ge 0 \\
{\rm (B)}
\nts&:&\nts x^2 + y^2 = 1 - z^2 \le 1,\es z \le 0
\end{eqnarray*}
\item \underline{Two donuts $\rightleftharpoons$ lens space
$L(p,\ts q)$}~\cite{lensSpace}
%
%
\begin{figure}[h]
\vspace{0mm}
\begin{center}
\setlength{\unitlength}{1cm}
\begin{picture}(8.0,6.8)(0,0)
\put(1.0,0.0){\includegraphics[width=6.0cm]{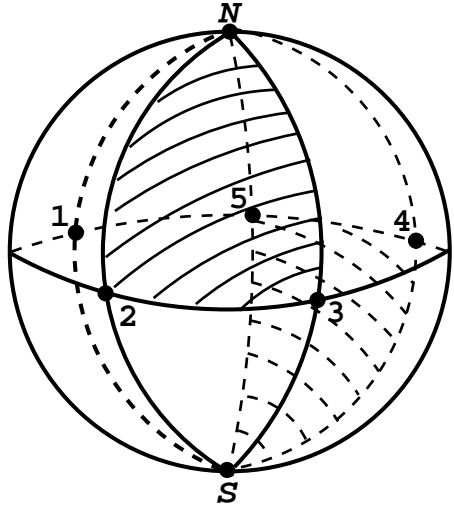}}
\end{picture}
\end{center}
\vspace*{-0.8cm}
\caption{Lens space $L(5,\ts 2)$. There are 5-sections each in upper and lower 
hemisphere. We identify the surface of the upper $n$~(modulo 5)-th section 
with the surface of the lower $(n+2)$~(modulo 5)-th section.}
\end{figure}
\end{enumerate}
%
\subsection{Geometry}
Thurston conjecture:
\begin{quote}
{\sl ``Any geometry can be brought into the following 8 geometries by a
diffeo(homeo)morphism: \\
$H^3,\ts S^3,\ts E^3,\ts S^2 \times S^1,\ts H^2 \times S^1,\ts
Sol,\ts, Nil,$ and $SL(2,R)$ ''}
\end{quote}
There is no need to integrate over $g_{ij}$ in eq.~(\ref{eq:clsdMembrnVacAmp}).
This situation is similar to the string case where $g_{ij}$-integration is 
reduced to a finite number of moduli integrations for a given topology. This 
is a very encouraging situation towards the proof of the renormalizability.

  Unitarity, however, seems to be an outstanding problem in this approach. We 
are not sure yet how this problem can be tackled. One way could be to go to a 
limit where one can apply the duality argument to utilize the string result.
%
\section{Low-energy effective action}\label{sect:lowEnEffAct}
\spc 11-dimensional supergravity~\cite{SUGRA_11dim} as it stands has a  huge 
``vacuum degeneracy'', which may be related to the supersymmetry of the theory.
This implies that we should look for a non-generic M-theory which admits only
restricted $g_{\mu\nu}$ or $A_{\mu\nu\rho\kappa}$ leading to the unique vacuum 
(mirror symmetry\/(?)~\cite{mirrorSymmetry}). The uniqueness of the vacuum
does not happen in the case of string theory where the conformal invariance 
leads to the Einstein Lagrangian and nothing more.

\noindent{Question:}
Does the phenomenologically correct vacuum exist among the possible vacua\/?
If this is the case, we may work backwards to find a principle to get this 
vacuum as a unique solution. Many authors have worked~\cite{stringVac} on the 
scheme in which 
\begin{eqnarray*}
11 \nts&=&\nts \underbrace{4}_{\mbox{\sp Minkowski}} \quad + 
\underbrace{7}_{\mbox{\sp compactified manifold} \atop
\mbox{\sp of $G_2$ holonomy}} \ts, \\
 7 \nts&=&\nts \underbrace{4}_{\mbox{\sp $K_3$ manifold}} + \quad\quad 3 \ts.
\end{eqnarray*}
The first is the consequence of supersymmetry and the second is the consequence
of mirror symmetry. The four-dimensional space ($K_3$ manifold) has a 
singularity group (monodromy group) to produce a given set of gauge particles.
This gauge symmetry is localized in the 3 space, as was first shown by Katz 
and Vafa~\cite{KatzVafa}, and extensively discussed later by Acharya and 
Witten~\cite{AcharyaWitten}. The supersymmetry may also be localized in the
3 space, although there is no rigorous proof on this. The proof could proceed
similarly to the case of gauge symmetry.
\vspace*{0.3cm}

\noindent{\underline{Assumption about the vacuum}} \\
\spn For phenomenological purposes let us assume the followings:
\begin{enumerate}
\setlength{\itemsep}{-2pt}
\item The 4 space ($K_3$ manifold) gives $SO(10)$ group symmetry.
\item The 3 space is a lens space $L(4,2)$ with certain discrete symmetries.
\end{enumerate}
We note that this kind of model will have a dual formulation~\cite{dualForm}
in the string models which utilize the D-branes.
\vspace{0.3cm}

\begin{figure}
\begin{center}
\setlength{\unitlength}{1cm}
\begin{picture}(6.0,6.0)(0,0)
\put(0.0,0.0){\includegraphics[width=6.0cm]{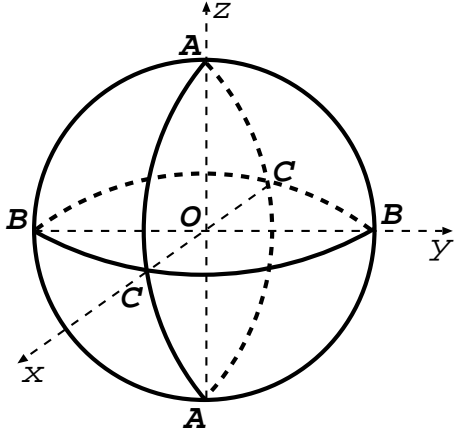}}
\end{picture}
\end{center}
\vspace*{-0.8cm}
\caption{Lens space $L(4,\ts 2)$.}
\label{fig:lensSpc42}
\end{figure}

\noindent{\underline{$\spadesuit$~Lens space $L(4,2)$}} \\
\spn We use $L(4,2)$ as is shown in Fig.~\ref{fig:lensSpc42}. The idea
is to obtain 3+1 fixed points where 3 corresponds to three generations and 1
to the hidden sector.
%
%
For this purpose we impose the following discrete symmetries to the
localizing equation of the symmetry.
\begin{eqnarray*}
\left\{
\begin{array}{rl}
(1) & y \to - y \quad {\rm and} \\
    & \nts\nts (y=0 \mbox{ surface and } B \mbox{ are invariant}) \\
(2) & z \to - z \\
(3) & x \to - x
\end{array}\right.
\label{eq:LSpc42Sym}
\end{eqnarray*}
$A,B,C,O$ in Fig.~\ref{fig:lensSpc42} are the only fixed points under the
combined symmetry of the above (1), (2) and (3). We assume that the theory is 
invariant under these transformations and the symmetry is localized in these 
points in the following way: The points $A,\ts B,\ts {\rm and}\es C$ each have 
a localized $SO(10)$ symmetry and the point $O$ has the $U(1) \times U(1)$
symmetry. The discrete symmetry (1), (2) or (3) is the artifact of low-energy 
theory, just like all the other global symmetries, such as P, CP, baryon 
number,
lepton number and so on. This implies that we are working on the theory not at 
$10^{19}~{\rm GeV}$, but rather at $10^{16}~{\rm GeV}$ or lower.
\vspace*{3mm}
%
\section{Explicit model}\label{sect:expModel}
\spc We now construct an explicit model based on the following multiplets:
\begin{itemize}
\setlength{\itemsep}{-2pt}
\item[(1)] Chiral multiplets:
\vspace*{-0.1cm}
\begin{eqnarray*}
\begin{array}{l}
Q^{(i)}_{16} \es (i=1,2,3) \quad 
1 \leftrightarrow A, \ts 2 \leftrightarrow B,\ts 3 \leftrightarrow C \\
H^{(1)}_{10} \mbox{ only at } A \\
H^{(i)}_{45} \es (i=1,2,3)
\end{array}
\end{eqnarray*}
Let us call these the {\em physical sector chiral multiplets}. $Q^{(i)}_{16}$
is the quark-lepton of the $i$-th generation and $H^{(1)}_{10}$ is the 
usual Higgs. Note that it couples only to the first generation (the 3rd 
generation in the usual terminology). We could have introduced $H^{(2)}_{10}$
and $H^{(3)}_{10}$ and forbid the coupling by the discrete symmetry, but
we have no compelling reason to do so.
\vspace*{0.2cm}
\item[(2)] Gauge fields:
\vspace*{-0.2cm}
\begin{eqnarray*}
SO(10) 
&:&\nts\nts V^{(i)}_{45} \es (i=1,2,3) \\
U(1)_y \times U(1)_z
\nts\nts\nts &:&\nts\nts V_y \es {\rm and} \es V_z
\end{eqnarray*}
\item[(3)] ``Hidden'' sector~(point $O$) $U(1)_y \times U(1)_z$ chiral 
multiplets:
\vspace*{-0.1cm}
\begin{eqnarray*}
\begin{array}{ll}
H(y_i),\ts H(-y_i) & (i=1,2,3) \\
H(z),\ts H(-z) \ts,
\end{array}
\end{eqnarray*}
where $y_i$ and $z$ stand for $U(1)_y$ and $U(1)_z$ charges, respectively.
The interpretation of these multiplets is that there could be others, but we 
identified those which couple to the physical sector through messengers.
\item[(4)] Messenger sector multiplets:
\vspace*{-0.1cm}
\begin{eqnarray*}
\begin{array}{ll}
Q^{(i)}_{16}(y_i),\es Q^{(i)}_{\overline{16}}(-y_i) & (i=1,2,3) \\
\vspace*{-4mm} \\
H^{(1)}_{10}(z), \es H^{(1)}_{10}(-z)
\end{array}
\end{eqnarray*}
These transform non-trivially, both in $SO(10)_i$ and in either
$U(1)_z$ or $U(1)_y$.
\end{itemize}

\noindent{\underline{Discrete symmetry}} \\
\spn The chiral multiplets are supposed to transform in the following way
under the product transformation of $x \to -x,\ts y \to -y$ and $z \to -z$:
\begin{eqnarray*}
&& Q^{(i)}_{16} \to e^{i\alpha^{(i)}_{16}} Q^{(i)}_{16} \quad {\rm with} \es 
\alpha^{(i)}_{16} = -\pi/4 \ts, \\
&& H^{(1)}_{10} \to e^{i\alpha_{10}} H^{(1)}_{10} \quad {\rm with} \es
\alpha_{10} = \pi/2 \ts, \\
{\rm and} \nms
&&H^{(i)}_{45} \to H^{(i)}_{45} \ts.
\end{eqnarray*}
The hidden sector multiplets transform trivially:
\vspace*{-1mm}
\begin{eqnarray*}
&& H(y_i) \to H(y_i) \ts, \\
&& H(-y_i) \to H(-y_i) \ts, \\
&& H(z) \to H(z) \ts, \\
{\rm and} \nms
&& H(-z) \to H(-z) \ts.
\end{eqnarray*}
The messenger sector multiplets transform in the following way:
\vspace*{-1mm}
\begin{eqnarray*}
&& Q^{(i)}_{\overline{16}}(-y_i) \to
e^{i\alpha^{(i)}_{\overline{16}}(-y_i)} 
Q^{(i)}_{\overline{16}}(-y_i) \quad {\rm with} \es 
\alpha^{(i)}_{\overline{16}}(-y_i) = \pi / 4 \ts, \\
&& Q^{(i)}_{\overline{16}}(y_i) \to
e^{i\alpha^{(i)}_{\overline{16}}(y_i)} 
Q^{(i)}_{\overline{16}}(y_i) \quad {\rm with} \es 
\alpha^{(i)}_{\overline{16}}(y_i) = -\pi / 4 \ts, \\
&& H_{10}(z) \to
e^{i\alpha_{10}(z)} H_{10}(z)  \quad {\rm with} \es
\alpha_{10}(z) = -\pi/2 \ts, \\
{\rm and} \nms
&& H_{10}(-z) \to
e^{i\alpha_{10}(-z)} H_{10}(-z) \quad {\rm with} \es
\alpha_{10}(-z) = +\pi/2 \ts.
\end{eqnarray*}
The gauge multiplets are all invariant under these transformations.
\vspace*{3mm}

\noindent{\underline{Superpotential}} \\
\spn We now write down all the possible terms in the superpotential.
\vspace*{-1mm}
\begin{eqnarray}
W 
\nms&=&\nms
g Q^{(1)}_{16} Q^{(1)}_{16} H^{(1)}_{10}  \nonumber \\
\nms&+&\nms g_i Q^{(i)}_{16} H(y_i) Q^{(i)}_{\overline{16}}(-y_i) 
            + h H^{(1)}_{10} H(-z)H^{(1)}_{10}(z) \nonumber \\
\nms&+&\nms f_i Q^{(i)}_{16}(y_i) H^{(i)}_{45} Q^{(i)}_{\overline{16}}(-y_i)
       + f_{10} H^{(1)}_{10}(z) H^{(1)}_{45} H^{(1)}_{10}(-z) \nonumber \\
\nms&+&\nms m_i Q^{(i)}_{16}(y_i) Q^{(i)}_{\overline{16}}(-y_i)
       + m_{10} H^{(1)}_{10}(z) H^{(1)}_{10}(-z) \nonumber \\
\nms&+&\nms M_i H(y_i) H(-y_i) + M_0 H(z) H(-z) \es.
\label{eq:superpotential}
\end{eqnarray}
Here the gauge group indices are not written explicitly. The first term is the 
only coupling among the physical sector multiplets. The second line describes 
the coupling of the physical multiplets $Q^{(i)}_{16}$ or $H^{(1)}_{10}$ to the
hidden sector multiplets $H(y_i)$ or $H(-z)$ through the corresponding 
messenger sector multiplets $Q^{(i)}_{\overline{16}}(-y_i)$ or 
$H^{(1)}_{10}(z)$, respectively. The third line is the coupling of 
$H^{(i)}_{45}$ to the messenger multiplets. The fourth line is the mass term
for the messenger multiplets and the fifth line is the mass term for the
hidden multiplets. We note the following:
\vspace*{-1mm}
\begin{enumerate}
\setlength{\itemsep}{-2pt}
\item The second and the third generations of quarks and leptons do not couple
to the Higgs multiplet $H^{(1)}_{10}$directly.
\item No mass term is possible for $H^{(1)}_{10}$.
\item $H^{(i)}_{45}$ couples only to $Q^{(i)}_{16}(y_i)Q^{(i)}_{16}(-y_i)$
and $H^{(1)}_{10}(z)H^{(1)}_{10}(-z)$ for $i=1$.
\item Discrete symmetry will not be used for the explanation of
doublet-triplet splitting. This will be explained later.
\item All the coupling constants can be calculated in principle in terms of
membrane instantons~\cite{BecBecSt}. The most important property we expect from
this computation, which we assume here, is the smallness of $m_{10}$ and $M$,
which should be of the order of TeV, compared to the other masses which are of 
the scale of $10^{16}~{\rm GeV}$ or higher. This is needed to explain
the doublet-triplet splitting in our scheme.
\end{enumerate}
%
\section{Broken symmetry and hierarchy}\label{sect:symbrkHrchy}
\spn We have to consider two kinds of symmetry breaking:
\begin{itemize}
\vspace*{-2mm}
\setlength{\itemsep}{-2pt}
\item[(1)] Supersymmetry breaking, and
\item[(2)] $SO(10) \times SO(10) \times SO(10) 
\to SU(3) \times SU(2) \times U(1) \ts$.
\end{itemize}
\vspace*{-3mm}
%
%
\subsection{Supersymmetry breaking}
\spn We assume that $D$-terms for $U(1)_z$ and $U(1)_y$ break the 
supersymmetry. The other motivation to include the $D$-terms in our model is to
circumvent the non-renormalization theorem which would have resulted
in~\cite{nonRenTheorem}
\begin{eqnarray*}
\langle Q^{(i)}_{16}(y_i) Q^{(j)}_{\overline{16}}(-y_j) \rangle 
\propto \delta_{ij} \es.
\end{eqnarray*}
This would have made it impossible to break $\{ SO(10) \}^3$ down to $SO(10)$ 
through the mechanism of the pair condensation of 
$Q^{(i)}_{16}(y_i) Q^{(j)}_{\overline{16}}(-y_j)$.
The non-renormalization theorem also would have lead to
\begin{eqnarray*}
\langle H_{10}(z) H_{10}(-z) \rangle = 0 \es.
\end{eqnarray*}
This would have made Higgs mass vanish even in the non-perturbative calculation
in our approach. The existence of $D$-terms is a simple way to circumvent this 
non-renormalization theorem.

  Explicitly, we get for the $D$-term of the form 
$\kappa_y D_y + \kappa_z D_z$ with
\vspace*{2mm}
\begin{eqnarray*}
\frac{\ds |M_i|^2}{\ds g_y y_i} < \kappa_y < 
\frac{\ds |m_i|^2}{\ds g_y y_i} \ts, \quad
\frac{\ds |M_0|^2}{\ds g_z z} < \kappa_z < 
\frac{\ds |m_{10}|^2}{\ds g_z z} \es,
\end{eqnarray*}
\vspace*{-3mm}
we get
\vspace*{-2mm}
\begin{eqnarray*}
\langle H(z) \rangle_F =
M_0 \langle H(-z) \rangle_A \ne 0 \ts, \quad{\rm and}\quad
\langle H(y_i) \rangle_F =
M_i \langle H(-y_i) \rangle_A \ne 0 \ts.
\end{eqnarray*}
This leads to the following consequences:
\begin{enumerate}
\setlength{\itemsep}{-2pt}
\item The $g_i Q^{(i)}_{16}\langle H(y_i) \rangle_F Q^{(i)}_{\overline{16}}
(-y_i)$ term in the superpotential gives a squark mass proportional to
$g_i \frac{\ds \langle H(y_i) \rangle_F^2}{\ds m_i}$. This means that all 
squark and slepton masses are of the see-saw type.
\item The $h H^{(1)}_{10} \langle H(-z) \rangle_A H^{(1)}_{10}(z)$ term
gives a Higgsino mass matrix of the form:
\begin{eqnarray*}
\left(
\begin{array}{cc}
0 & \langle H(-z) \rangle_A \\
\langle H(-z) \rangle_A & m_{10}
\end{array}
\right) \es.
\end{eqnarray*}
This is not necessarily a see-saw type because we assume $m_{10}$ is small 
(TeV). The gaugino mass of $V^{(i)}_{45}$
is given by the diagram in Fig.~\ref{fig:gauginoMass}.
%
%
\begin{figure}[h]
\vspace{0mm}
\begin{center}
\setlength{\unitlength}{1cm}
\begin{picture}(12.0,4.0)(0,0)
\put(1.0,0.0){\includegraphics[width=10.0cm]{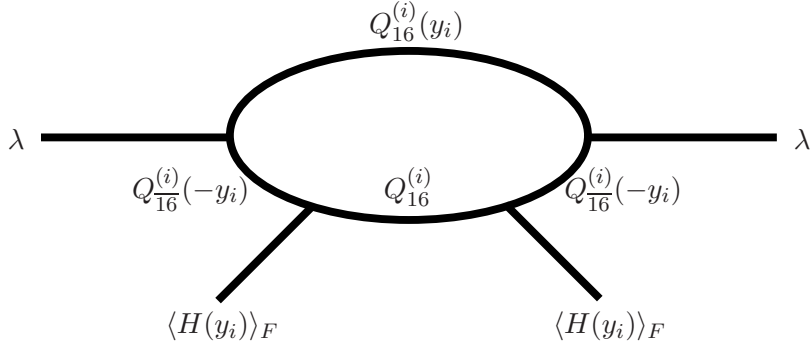}}
\put(0.7,2.15){$\lambda$}
\put(11.15,2.15){$\lambda$}
\put(5.5,3.7){$Q^{(i)}_{16}(y_i)$}
\put(5.7,1.5){$Q^{(i)}_{16}$}
\put(2.35,1.5){$Q^{(i)}_{\overline{16}}(-y_i)$}
\put(8.1,1.5){$Q^{(i)}_{\overline{16}}(-y_i)$}
\put(2.8,-0.3){$\langle H(y_i) \rangle_F$}
\put(7.95,-0.3){$\langle H(y_i) \rangle_F$}
\end{picture}
\end{center}
\vspace*{-0.3cm}
\caption{Gaugino masses.}
\label{fig:gauginoMass}
\end{figure}

\spn Other physical sector masses can be calculated in the following way:
\begin{itemize}
\setlength{\itemsep}{-2pt}
\item[1)] Higgs masses can be calculated using the diagram shown in
Fig.~\ref{fig:higgsMass} with appropriate supersymmetry breaking
and the symmetry breaking taken into account to evade the non-renormalization 
theorem. As we will see later, the internal masses are of the order TeV for the
doublet component of $H^{(1)}_{10}$ and of the order $10^{16}$~GeV for the 
triplet, giving the doublet-triplet splitting.
%
%
\begin{figure}[h]
\vspace{0mm}
\begin{center}
\setlength{\unitlength}{1cm}
\begin{picture}(10,2.6)(0,0)
\put(1.0,0.0){\includegraphics[width=8.0cm]{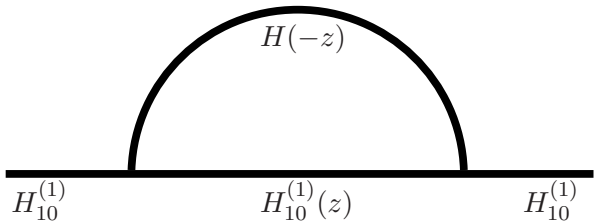}}
\put(4.5,1.8){$H(-z)$}
\put(1.2,-0.4){$H^{(1)}_{10}$}
\put(4.5,-0.4){$H^{(1)}_{10}(z)$}
\put(8.0,-0.4){$H^{(1)}_{10}$}
\end{picture}
\end{center}
\vspace*{-0.15cm}
\caption{Higgs masses.}
\label{fig:higgsMass}
\end{figure}
\item[2)] Fig.~\ref{fig:nu_RMass} gives only the $\bar{\nu}^c_{R} \nu_{R}$
mass term because we do not violate the $SU(3) \times SU(2) \times U(1)$ till
the vacuum value of Higgs doublet is taken into account. The masses of quarks 
and leptons are the subject of section~\ref{sect:QrkLptnMtrx}. 
%
%
\begin{figure}
\vspace{0mm}
\begin{center}
\setlength{\unitlength}{1cm}
\begin{picture}(10,2.6)(0,0)
\put(1.0,0.0){\includegraphics[width=8.0cm]{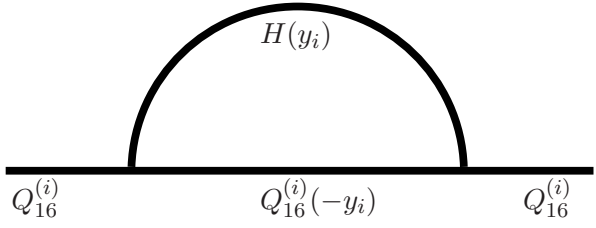}}
\put(4.5,1.8){$H(y_i)$}
\put(1.2,-0.4){$Q^{(i)}_{16}$}
\put(4.5,-0.4){$Q^{(i)}_{16}(-y_i)$}
\put(8.0,-0.4){$Q^{(i)}_{16}$}
\end{picture}
\end{center}
\vspace*{-1.5mm}
\caption{$\nu_{R}$ mass diagram.}
\label{fig:nu_RMass}
\end{figure}
\end{itemize}
\end{enumerate}
%
\subsection{Gauge symmetry breaking}
\spc The $\{ SO(10) \}^3$ breakdown to $SO(10)$ will be achieved by having
\begin{eqnarray*}
\langle Q^{(i)}_{16}(y_i) Q^{(j)}_{\overline{16}}(-y_j) \rangle = m_{ij} \es.
\end{eqnarray*}
The $D$-term can break the supersymmetry and it will also break the powerful 
non-renormalization theorem. In a way, this is the origin of ``flavor 
physics''. For this purpose the non-perturbative effect must be considered
in the sense of Nambu-Jona-Lasinio~(NJL) or Bardeen-Cooper-Shrieffer~(BCS).
Self-consistent mass generations \`a la NJL or BCS can be expressed by the
following equation:
\begin{eqnarray*}
m_{ij} 
\nms&=&\nms m_i \delta_{ij} + \Sigma(m_{ij}) \ts, \quad 
\mbox{which comes from}\ts, \\
\Gamma(\varphi) 
\nms&=&\nms S(\varphi) + \frac{\ds 1}{\ds 2}
{\rm tr} \log \left[ \frac{\ds \delta^2 S}
{\ds \delta \varphi_i \delta \varphi_j} \Big/ \frac{\ds \delta^2 S}
{\ds \delta \varphi_i \delta \varphi_j}\Big|_{\varphi=0}
\right] \ts,
\end{eqnarray*}
where $\Gamma$ is the one-particle irreducible potential and $\varphi$ 
stands for the generic field.

%
%
\begin{figure}[h]
\vspace{-0.0cm}
\begin{center}
\setlength{\unitlength}{1cm}
\begin{picture}(12.0,1.6)(0,0)
\put(0.0,0){\includegraphics[width=12.0cm]{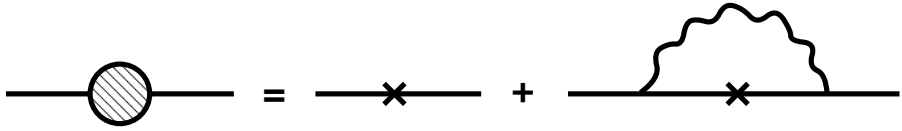}}
\end{picture}
\end{center}
\vspace*{-8mm}
\caption{Self-consistent mass.}
\label{fig:massGen}
\end{figure}

\begin{figure}[h]
\vspace{-0.0cm}
\begin{center}
\setlength{\unitlength}{1cm}
\begin{picture}(5.0,4.7)(0,0)
\put(0.0,0.0){\includegraphics[width=5.0cm]{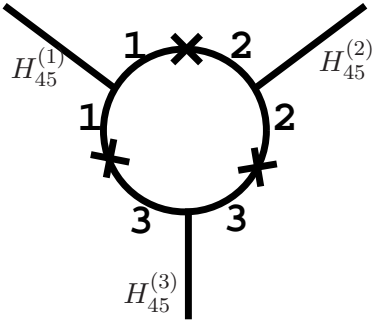}}
\put(0.2,3.3){$H^{(1)}_{45}$}
\put(4.3,3.4){$H^{(2)}_{45}$}
\put(1.7,0.3){$H^{(3)}_{45}$}
\end{picture}
\end{center}
\vspace*{-0.7cm}
\caption{Cubic term.}
\label{fig:cubicTerm}
\end{figure}

  It is interesting to note that when we break $SO(10)$ down to $SU(3) \times
SU(2) \times U(1)$, we need an effective cubic term for the $H_{45}$ 
potential. We have
\begin{eqnarray*}
{\rm tr} \left( 
H^{(1)}_{45} H^{(2)}_{45} H^{(3)}_{45} \right) \ne 0 \es,
\end{eqnarray*}
due to the diagram in Fig.~\ref{fig:cubicTerm}. This is very different from
the single $SO(10)$ case where we need to add other multiplets of 
higher representations~\cite{higherRep}.
%
%

\begin{figure}[h]
\vspace{0.0cm}
\begin{center}
\setlength{\unitlength}{1cm}
\begin{picture}(12.0,1.8)(0,0)
\put(0.0,0.3){\includegraphics[width=12.0cm]{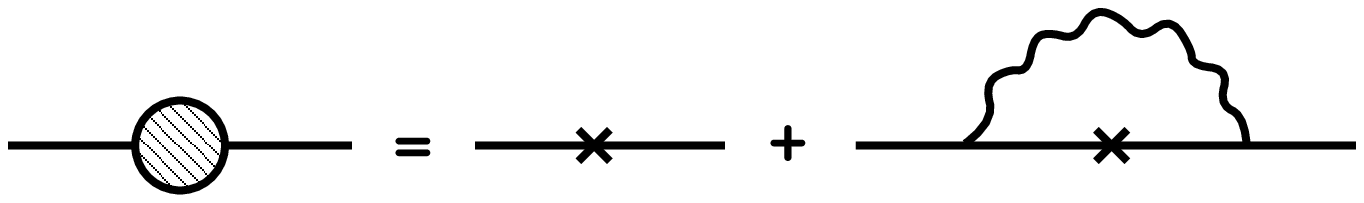}}
\put(1.3,0.0){$m_{ij}$}
\put(3.5,0.0){$= \quad\quad\ts m_i \delta_{ij}$}
\put(8.4,0.3){$g_{i}$}
\put(10.9,0.3){$g_{j}$}
\end{picture}
\end{center}
\vspace*{-0.7cm}
\caption{Simplified self-consistent equation.}
\label{eq:smplfdSlfCnstnt}
\end{figure}

   In our case, the self-consistent mass equation (gap equation) 
takes the form
%
\vspace*{-1mm}
\begin{eqnarray*}
\Gamma_{\rm mass}(\varphi) = S(\varphi) - \frac{\ds 1}{\ds 2} {\rm Tr} \int
d\Omega d\Omega^{\prime} \nms\nms
& &\nms\nts \Bigl[ K^{-1}(\Omega - \Omega^{\prime})_{\varphi\varphi} 
a_{\varphi V}(\Omega^{\prime}) K^{-1}(\Omega^{\prime} - \Omega)_{VV} 
a_{V \varphi}(\Omega) \\
&+&\nts\nts  K^{-1}(\Omega - \Omega^{\prime})_{\varphi V} 
a_{V \varphi}(\Omega^{\prime}) K^{-1}(\Omega^{\prime} - \Omega)_{\varphi V} 
a_{V \varphi}(\Omega) \Bigr] \es,
\end{eqnarray*}
\vspace*{-3mm}
where\footnote{The notations here come from J.~Wess and J.~Bagger, {\it 
``Supersymmetry and Supergravity''}, Princeton University Press, 1992.} 
\vspace*{-2mm}
\begin{eqnarray*}
d\Omega \equiv dx d^2\theta d^2\bar{\theta},\es
\frac{\ds \delta^2 S}{\ds \delta \varphi_i \delta \varphi_j} \equiv
\underbrace{K_{ij}}_{L_0} + \underbrace{a_{ij}}_{L_{\rm int}} \ts.
\end{eqnarray*}
For example,
\begin{eqnarray*}
K_{V \varphi} =
\left(
\begin{array}{ccc}
\vspace*{2mm}
\nms - \Box P_T + \frac{\ds 1}{\ds 2} m_V^2 - \xi (P_1 + P_2) \Box
& -2gYA^{\dagger}(y) & 0 \\
\vspace*{2mm}
0 & -\frac{\ds m}{\ds 4 \Box}DD & 1+gY D\theta^2\bar{\theta}^2 \\
\vspace*{2mm}
-2gYA(-y) & 1-gYD\theta^2\bar{\theta}^2 
& -\frac{\ds m}{\ds 4 \Box} \bar{D}\bar{D}
\end{array}
\right) \es, \quad\quad {\rm etc.}
\end{eqnarray*}
  Rather than tackling this complicated equation here, I would like to solve
some prototype equations where supersymmetry is completely neglected just to
illustrate the basic idea involved.
The equation is shown in Fig.~\ref{eq:smplfdSlfCnstnt}.
%
\begin{itemize}
\item[(1)] $H_{10}^{(1)}$ case \\
\spn This is a single channel case and is very simple:
\vspace*{-1mm}
\begin{eqnarray*}
m 
\nms&=&\nms m_0 + g^2 m \log \frac{\ds \Lambda}{\ds m} \ts, \\
m_0 \nts
\nms&=&\nms m \left( 1 - g^2 \log \frac{\ds \Lambda}{\ds m} \right)\ts, 
\es{\rm or} \es
m = \frac{\ds m_0}{\ds 1 - g^2 \log \frac{\ds \Lambda}{\ds m}} \ts.
\end{eqnarray*}
\vspace*{-8mm}

\noindent{The} solution is illustrated in Fig.~\ref{eq:massSplttng}.
For $m_0= 0$, we get
\begin{eqnarray*}
m = 0 \es{\rm or,}\es m = \Lambda e^{-1/g^2} \es.
\end{eqnarray*}
The point is that the scale $\Lambda$ of the broken symmetry solution $m = 
\Lambda e^{-1/g^2}$ has nothing to do with the scale of the bare mass $m_0$.
The symmetry is combined with the diagram in Fig.~\ref{fig:higgsMass}. 
When the symmetry is broken down to $SU(3) \times SU(2) \times U(1)$, we may 
arrange the coupling in such a way that the doublet corresponds to the 
unbroken solution and the triplet corresponds to the broken solution 
($m \simeq \Lambda e^{-1/g^2}$).\footnote{For $\Lambda \sim 10^{19}~{\rm GeV}$
and $m \sim 10^{16}~{\rm GeV}$, we have $g^2 \sim 0.1$.}.
This, combined with the diagram in 
Fig.~\ref{fig:higgsMass}, is our mechanism of the doublet-triplet splitting.
\begin{figure}[h]
\begin{center}
\setlength{\unitlength}{1cm}
\begin{picture}(8.5,5.2)(0,0)
\put(0.0,0.3){\includegraphics[width=8.0cm]{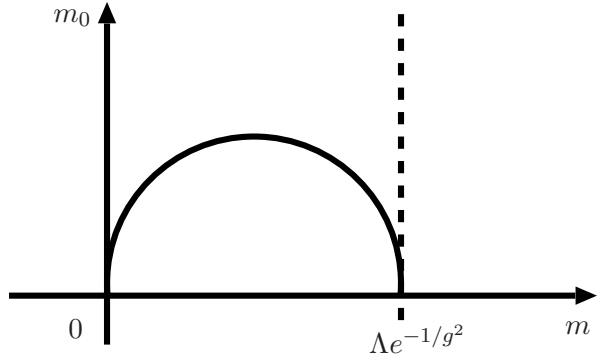}}
\put(0.7,4.7){$m_0$}
\put(0.9,0.5){$0$}
\put(4.9,0.3){$\Lambda e^{-1/g^2}$}
\put(7.5,0.55){$m$}
\end{picture}
\end{center}
\vspace*{-1.2cm}
\caption{One channel solution.}
\label{eq:massSplttng}
\end{figure}
\vspace*{-2mm}
\item[(2)] $Q^{(i)}_{16}(y_i)$ case \\
\spn The most important question in this case is whether we can have a solution
with the off-diagonal elements. We have
\begin{eqnarray*}
m_{ij} = m_0 \delta_{ij} + g_i 
\left( m \log \frac{\ds \Lambda}{\ds m} \right)_{ij} g_j
\quad\quad (i = 1,\ts, 2,\ts 3) \es,
\end{eqnarray*}
where no summation over $i,\ts j$ is implied.
We consider the case where,
\begin{eqnarray*}
m_{ij} = m_0 \delta_{ij} + \overline{m}_{ij} \ts, \quad
|\overline{m}_{ij}| \le |m_0| \ts.
\end{eqnarray*}
We use the notation
\begin{eqnarray*}
x_{ij} = \frac{\ds \overline{m}_{ij}}{\ds m_0} \ts, \quad
\lambda \equiv \log\frac{\ds \Lambda}{\ds m_0} \ts, \quad \mbox{and}
\quad |g_1| > |g_2| > |g_3| \ts, \quad \lambda \gg 1 \es.
\end{eqnarray*}
Then, the equation becomes
\begin{eqnarray*}
x_{ij} = \lambda g_i g_j \delta_{ij} + (\lambda - 1) g_i g_j x_{ij} -
\frac{\ds 1}{\ds 2} g_i g_j \left( x^2 \right)_{ij} \es.
\end{eqnarray*}
We get, as the approximate solution,
\begin{eqnarray*}
x_{11} \simeq 1/g_1 \es.
\end{eqnarray*}
\vspace*{-3mm}
Assuming
\begin{eqnarray*}
g_2 
\ge \frac{\ds 1}{\ds (\lambda - 1)g_1 + 1/2} \ts, \quad 
g_3 \le \frac{\ds 1}{\ds (\lambda - 1)g_1 + 1/2} \ts, 
\end{eqnarray*}
we get
\vspace*{-3mm}
\begin{eqnarray*}
x_{22} \simeq \lambda g_2^2 \ts, \quad x_{33} \simeq \lambda g_3^2 \ts,
\end{eqnarray*}
\vspace*{-3mm}
and
\vspace*{-1mm}
\begin{eqnarray*}
x_{12}^2 
\nms&=&\nms \frac{\ds 4}{\ds g_1 g_2 g_3^2} \left\{ 
1-(\lambda - 1)g_1g_3 + \frac{1}{2} g_1g_3 \left( x_{11}+x_{33} \right)
\right\} \ts, \\
x_{13}^2 
\nms&=&\nms \frac{\ds 4}{\ds g_1 g_3 g_2^2} \left\{ 
1-(\lambda - 1)g_1g_2 + \frac{1}{2} g_1g_2 \left( x_{11}+x_{22} \right)
\right\} \ts, \\
x_{23}^2 
\nms&=&\nms x_{12}^2 x_{13}^2 \cdot \frac{1}{2} g_2^2 g_3^2 \ts, 
\end{eqnarray*}
where $x_{12}$ is real, and $x_{13}$ and $x_{23}$ are imaginary, implying a
contribution to the CP violation. We see that the solution with the 
off-diagonal element exists and at least a part of CP violation may come from
the phase of the BCS-NJL order parameter.
\end{itemize}
%
\section{Quark-lepton mass matrix and mixing matrix calculation \\ 
{\small (This part of the work is in collaboration with 
S.~Pakvasa)}}\label{sect:QrkLptnMtrx}
\begin{figure}[h]
\begin{center}
\setlength{\unitlength}{1cm}
\begin{picture}(5.0,5.0)(0,0)
\put(0.0,0.0){\includegraphics[width=5.0cm]{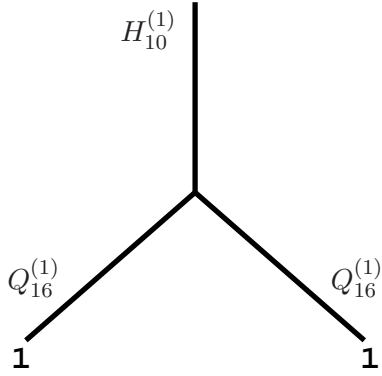}}
\put(1.5,4.5){$H^{(1)}_{10}$}
\put(0.0,1.2){$Q^{(1)}_{16}$}
\put(4.3,1.2){$Q^{(1)}_{16}$}
\end{picture}
\end{center}
\vspace*{-1.0cm}
\caption{Higgs coupling.}
\label{fig:gQQH}
\end{figure}

\spc The $gQ^{(1)}_{16}Q^{(1)}_{16}H^{(1)}_{10}$ as in Fig.~\ref{fig:gQQH}
is the only Higgs coupling in our model.
\vspace*{2mm}

\begin{figure}[h]
\begin{center}
\setlength{\unitlength}{1cm}
\begin{picture}(12.0,5.6)(0,0)
\put(0.0,0.0){\includegraphics[width=12.0cm]{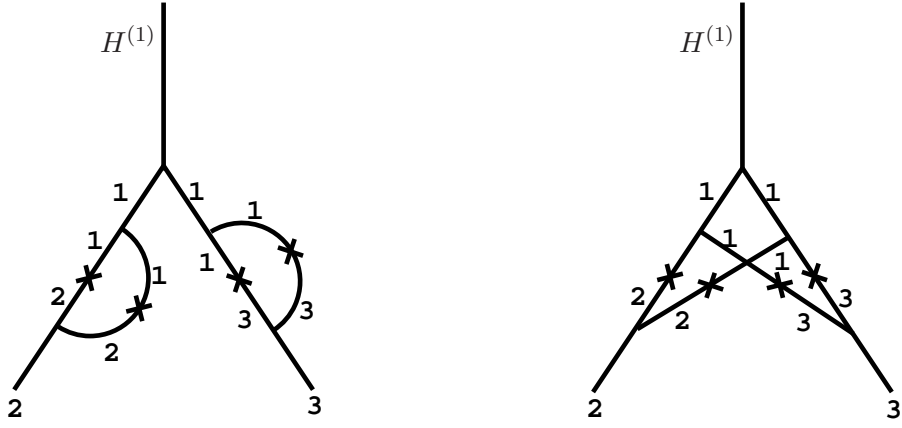}}
\put(1.3,5.0){$H^{(1)}$}
\put(9.0,5.0){$H^{(1)}$}
\end{picture}
\end{center}
\vspace*{-0.9cm}
\caption{(a) The second order couplings (external correction type).
(b) The second order couplings (vertex correction type).}
\label{fig:g2mQQHH}
\end{figure}

\begin{figure}[h]
\vspace{0cm}
\begin{center}
\setlength{\unitlength}{1cm}
\begin{picture}(3.8,4.9)(0,0)
\put(0.0,0.0){\includegraphics[width=3.8cm]{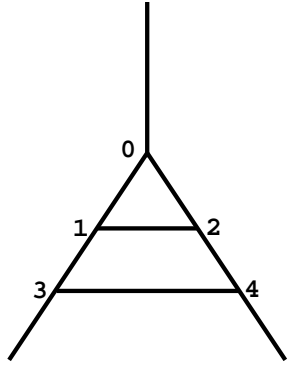}}
\end{picture}
\end{center}
\vspace*{-0.8cm}
\caption{Diagram of this kind does not exist.}
\label{fig:fbdnQQH}
\vspace*{0.3cm}
\end{figure}

  Other couplings can be calculated in the following way:
\begin{itemize}
\item[(1)] $Q^{(1)}_{16} Q^{(2)}_{16} H^{(1)}_{10}$ or $Q^{(1)}_{16} 
Q^{(3)}_{16} H^{(1)}_{10}$ can be calculated as in Fig.~\ref{fig:gQQHH}.
\begin{figure}[h]
\begin{center}
\setlength{\unitlength}{1cm}
\begin{picture}(8.0,7.0)(0,0)
\put(0.0,0.0){\includegraphics[width=7.0cm]{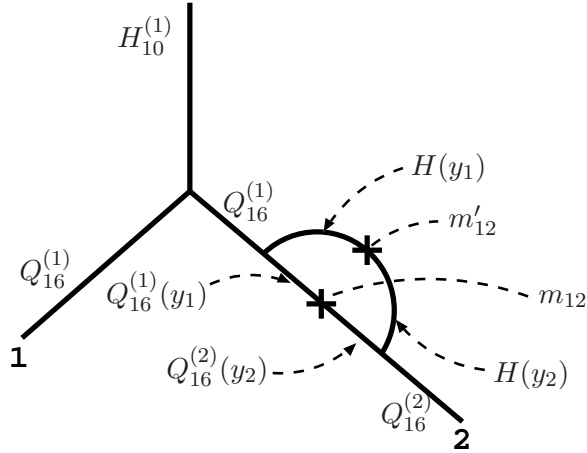}}
\put(1.5,5.5){$H^{(1)}_{10}$}
\put(0.2,2.4){$Q^{(1)}_{16}$}
\put(2.9,3.3){$Q^{(1)}_{16}$}
\put(5.4,3.8){$H(y_1)$}
\put(6.5,1.05){$H(y_2)$}
\put(1.35,2.1){$Q^{(1)}_{16}(y_1)$}
\put(2.15,1.1){$Q^{(2)}_{16}(y_2)$}
\put(5.0,0.5){$Q^{(2)}_{16}$}
\put(5.9,3.1){$m_{12}^{\prime}$}
\put(7.1,2.1){$m_{12}$}
\end{picture}
\end{center}
\vspace*{-1.0cm}
\caption{The first order couplings for $Q^{(1)}_{16}Q^{(2)}_{16}H^{(1)}_{10}$.
$Q^{(2)}_{16}$ is replaced by $Q^{(3)}_{16}$ to calculate the third generation
coupling.}
\label{fig:gQQHH}
\end{figure}
\item[(2)] $Q^{(2)}_{16} Q^{(3)}_{16} H^{(1)}_{10}$ can be calculated as in 
Fig.~\ref{fig:g2mQQHH}~(a) and (b). It is easy to show that the diagram of
Fig.~\ref{fig:fbdnQQH} does not exist.
\end{itemize}

  Mass matrices for up-type quarks $U$, down-type quarks $D$ and charged
leptons $L$, therefore, have the following form:
\begin{eqnarray*}
\left(
\begin{array}{ccc}
1 & c & d \\
a  & \alpha & \gamma \\
b & \beta & \delta
\end{array}\right) \es,
\end{eqnarray*}
with $a,b,c$ and $d$ being the first order and $\alpha,\beta,\gamma$ and
$\delta$ being the second order.

  The $\nu_R$ mass matrix has the form:
\begin{eqnarray*}
m_D =
\left(
\begin{array}{ccc}
\Delta_{11} & &  \\
  & \Delta_{22} & \\
 &  & \Delta_{33}
\end{array}
\right) \es,
\end{eqnarray*}
taking into account the smallness of the off-diagonal elements suggested by the
prototype computation in the previous section.

  After some computation we get the following KM matrix\footnote{The matrix 
here is in the grand unified scale. The renormalization effect can be 
calculated~\cite{BBO}.}:
\begin{eqnarray*}
\nms\nts\left(
\begin{array}{ccc}
1 & -(\chi^{\dagger}_-, V_{+d}) - \bar{\kappa}^d_{+-} + \bar{\kappa}^u_{+-} &
- \bar{\kappa}^d_{+0} + \bar{\kappa}^u_{+0} + \bar{\kappa}^d_{-0}
(\chi^{\dagger}_-, V_{+d}) \\
(\chi^{\dagger}_-, V_{+d}) + \kappa^d_{+-} - \kappa^u_{+-} &
1 - \frac{1}{2} |\kappa^u_{-0} - \kappa^d_{-0}|^2 &
- \bar{\kappa}^d_{-0} + \bar{\kappa}^u_{-0} \\
\kappa^d_{+0} - \kappa^u_{+0} - \kappa^u_{-0} (\chi^{\dagger}_-, V_{+d})
& \kappa^d_{-0} - \kappa^u_{-0} &
1 - \frac{1}{2} |\kappa^u_{-0} - \kappa^d_{-0}|^2
\end{array}
\right) \ts,
\end{eqnarray*}
where
\clearpage
\begin{eqnarray*}
&& V_+ =
\left(
\begin{array}{c}
1 \\ a \\ b
\end{array}
\right) ,\quad
V_- = \frac{\ds 1}{\ds \sqrt{|a|^2 + |b|^2}} \left(
\begin{array}{c}
|a|^2 + |b|^2 \\ -a \\ -b
\end{array} \right) , \hspace*{2cm} \\
{\rm and} 
&&
V_0 = \frac{\ds 1}{\ds \sqrt{|a|^2 + |b|^2}} \left(
\begin{array}{c}
0 \\ - \bar{b} \\ \bar{a}
\end{array}
\right) ,
\end{eqnarray*}
with the suffix $u$ or $d$ signifying the up- or down-type quark quantities.
We also have 
\begin{eqnarray*}
\chi_{\pm}
\nms&\equiv&\nms V_{\pm u} - V_{\pm d} \ts, \\
\kappa_{ik}
\nms&\equiv&\nms \frac{\ds 1}{\ds \lambda_i - \lambda_k}
\langle V_k | H_1 | V_i \rangle \ts, \quad H_1 = 
\left(
\begin{array}{ccc}
0 & c & d \\
a & 0 & 0 \\
b & 0 & 0
\end{array}
\right) , \hspace*{1cm}
\end{eqnarray*}
\begin{eqnarray*}
\lambda_+ 
\nms&=&\nms 1 + |a|^2 + |b|^2 + |c|^2 + |d|^2 \ts, \quad
\lambda_- = \frac{\ds (|a|^2 + |b|^2) (|c|^2 + |d|^2)}
{\ds 1 + |a|^2 + |b|^2 + |c|^2 + |d|^2} \ts, \\
{\rm and} \quad \lambda_0
\nms&=&\nms \frac{\ds 1}{\ds |c|^2 + |d|^2} \left\{ 
|a\alpha - c\gamma|^2 + |d\beta - c\delta|^2 \right\} 
+ \mbox{``second order''} \ts.
\end{eqnarray*}

  The neutrino mixing matrix
is also calculated to be:
\begin{eqnarray*}
\mbox{mixing matrix} 
\nms&=&\nms\left(
\begin{array}{ccc}
\vspace*{2mm}
\cos\theta\frac{\lambda^{\ast}_{+}}{|\lambda_{+}|} &
\cos\theta(a^{\ast}y - b^{\ast}x) &
\cos\theta(a^{\ast}x^{\ast} + b^{\ast}y^{\ast}) \\
\nms \sin\theta\frac{\lambda^{\ast}_{+}}{|\lambda_{+}|} &
\cos\theta\cot\theta(-a^{\ast}y + b^{\ast}x) &
\cos\theta\cot\theta(-a^{\ast}x^{\ast} - b^{\ast}y^{\ast}) \\
\frac{1}{N_+}\cot\theta(-by - ax) &
\cot\theta(-by - ax) &
\cot\theta(-bx^{\ast} - ay^{\ast})
\end{array}
\right)
\end{eqnarray*}
with the neutrino mass matrix being:
\begin{eqnarray*}
M \nms&=&\nms m_D 
\left(
\begin{array}{ccc}
\Delta_{11}^{-1} & & \\
 & \Delta_{22}^{-1} & \\
 & & \Delta_{33}^{-1}
\end{array}
\right) m_D^t 
\end{eqnarray*}
Here $\cos\theta = 1 / \sqrt{1+|a|^2+|b|^2}$. The $a$ and $b$ in this formula
correspond to mass matrix elements of the leptons. We also have for the 
neutrino parameters
\begin{eqnarray*}
\lambda_+ 
\nms&\simeq&\nms 1 + c^{\ast 2} + d^{\ast 2} \ts, \quad
x \simeq -\frac{\ds 1}{\ds N_0} \left\{ b + c\beta + d\delta \right\} \ts,\\
\lambda_- 
\nms&\simeq&\nms - \frac{\ds N_0^2}{\ds 1 + c^{2} + d^{2}} \ts, \quad
y \simeq \frac{\ds 1}{\ds N_0} \left\{ a + c\alpha + d\gamma \right\} \ts, \\
{\rm and} \quad N_+^2 
\nms&=&\nms \frac{\ds |\lambda_+|^2}{\ds N_0^2} + |x|^2 + |y|^2 \ts,
\end{eqnarray*}
with $a,\ts b,\ts {\rm etc}.$ corresponding to the left-handed neutrino mass 
matrix elements. Actually, in this formula for the neutrino parameters we 
replaced the original parameters in the following way: 
\begin{eqnarray*}
c^2 \Delta_{11}\Delta^{-1}_{22} \longrightarrow c^2 
&,& \quad d^2 \Delta_{11}\Delta^{-1}_{33} \longrightarrow d^2 \ts, \\
\alpha^2 \Delta^{-1}_{22} \longrightarrow \alpha^2
&,& \quad \beta^2 \Delta^{-1}_{22} \longrightarrow \beta^2 \ts, \\
\gamma^2 \Delta^{-1}_{3} \longrightarrow \gamma^2
&,& \ts{\rm and}\es \delta^2 \Delta^{-1}_{33} \longrightarrow \delta^2 \ts.
\end{eqnarray*}
This means that the contribution of the Majorana mass is to renormalize the
original $a,\ts b,\ts {\rm etc}.$. The actual computation of the $a,\ts b,\ts 
{\rm etc}.$ through the Feynman diagrams of Fig.~\ref{fig:gQQHH} and
Fig.~\ref{fig:g2mQQHH} is underway. We hope to present the results soon.
Interestingly enough, our expressions for the mixing angles, even without the
diagrams computation, already show some nice features: both quark and lepton
mixing matrices are consistent with experiments.
%
\section{Conclusions}\label{sect:conclusions}
\begin{itemize}
\setlength{\itemsep}{-1pt}
\item[(1)] The effort to formulate the M-theory as the renormalizable membrane 
theory is continued.
\item[(2)] The vacuum problem of the M-theory should be considered both from 
the theoretical and the phenomenological aspects. It is important to make sure 
that it contains a physically acceptable vacuum, thus requiring the necessity
of phenomenology.
\item[(3)]  A candidate vacuum is considered, which is a kind of lens space
with the discrete symmetries.
\item[(4)] $\{ SO(10) \}^3 \times (U(1) \times U(1))$ model is proposed
as the candidate for the phenomenology.
\item[(5)] Various hierarchy mechanisms are considered.
\vspace*{-2mm}
\begin{enumerate}
\setlength{\itemsep}{-4pt}
\item membrane instanton
\item see-saw
\item BCS-NJL order parameter
\end{enumerate}
\item[(6)] CKM and neutrino mixing are calculable as is explicitly 
demonstrated.
\item[(7)] New physics may show up in, for example, B-physics as is
shown in Fig.~\ref{fig:newPhy}.
%
%
\begin{figure}[h]
\begin{center}
\setlength{\unitlength}{1cm}
\begin{picture}(7,5.0)(0,0)
\put(0.0,0.0){\includegraphics[width=7cm]{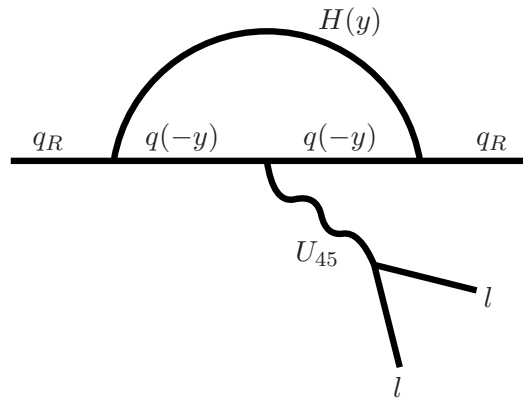}}
\put(0.4,3.05){$q_R$} \put(6.3,3.05){$q_R$}
\put(1.9,3.05){$q(-y)$} \put(4.0,3.05){$q(-y)$}
\put(4.2,4.6){$H(y)$}
\put(3.9,1.5){$U_{45}$}
\put(6.4,0.9){$l$} \put(5.2,-0.3){$l$}
\end{picture}
\end{center}
\vspace*{-0.6cm}
\caption{A possible new physics.}
\label{fig:newPhy}
\end{figure}
\end{itemize}
\vspace*{5mm}

\noindent{\Large\bf Acknowledgment} \\
\spn I would like to thank K.~Higashijima, S.~Pakvasa and X.~Tata for 
discussions. I am indebted the to Dai Ho Chun family for financial support 
during my stay at the University of Hawaii, and to all the UH High Energy 
Physics Group members for making my stay pleasant. I appreciate K.~Nakamura
and Fujihara Foundation for inviting me to the ``Seesaw Symposium''.
I am also thankful to H.~Hagura for making the OHP for the presentation and 
typing my barely readable manuscript.
%
%
%
%
\clearpage

%

\begin{thebibliography}{99}
\setlength{\itemsep}{-2pt}
\bibitem{renGroup} H.~Georgi and S.~L.~Glashow, 
{\sl Phys.~Rev.~Lett.}~{\bf 32} (1974) 438. \\
H.~Georgi, H.~R.~Quinn and S.~Weinberg, 
{\sl Phys.~Rev.~Lett.}~{\bf 33} (1974) 451.
\bibitem{GMRS} Gell-Mann, P.~Ramond and R.~Slansky, {\it ``Complex Spinors 
and Unified Theories'',\/} in {\sl Supergravity,\/}
eds.~P.~van Nieuwenhuizen and D.~Z.~Freedman, North Holland (1979) 315.
\bibitem{Yanagida} T.~Yanagida, {\it ``Horizontal Gauge Symmetry and Masses
of Neutrinos'',\/} in Proceedings of the Workshop on {\sl Unified Theories and
Baryon Number in the Universe}, eds.~O.~Sawada and A.~Sugamoto, 
KEK (1979) 95.
\bibitem{wittenG2} E.~Witten, {\it ``Deconstruction, $G_2$~Holonomy, and 
Doublet-Triplet Splitting'',\/} in 10th International Conference on 
{\sl Supersymmetry and Unification of Fundamental Interactions (SUSY02),\/} 
Hamburg, Germany, Jun.~17-23 2002, hep-ph/0201018.
\bibitem{Heegaard} Elementary explanation is given in
{\sl Tales of Poincar\'e Conjecture\/} by Tatsuo Honma (in Japanese),
Nihon Hyouron Sya (1985).
\bibitem{Thurston} W.~P.~Thurston, {\sl Three-Dimensional Geometry and
Topology,\/} Vol.~1, ed.~S.~Lovy, Princeton University Press (1977).
\bibitem{G_2holonomy} M.~F.~Atiyah and E.~Witten, {\it ``M-Theory Dynamics 
on A Manifold of $G_2$ Holonomy'',\/} 
{\sl Adv.~Theor.~Math.~Phys}.~{\bf 6}~(2003) 1, hep-th/0107177.
\bibitem{deconstruction} N.~Arkani-Hamed, A.~G.~Cohen and H.~Georgi,
Phys.~Rev.~Lett.~{\bf 86} (2001) 4757, hep-th/0104005.
\bibitem{explicitModel} $SU(5)^3$ was considered by S.~Rajapoot, 
{\sl Phys.~Rev.}~{\bf D24} (1981) 1890. \\
$SU(3)^3$ was considered by C.~D.~Carone and H.~Murayama,
{\sl Phys.~Rev.}~{\bf D52} (1995) 4159.
\bibitem{FayetIliopoulos} P.~Fayet and J.~Iliopoulos, 
{\sl Phys.~Lett.}~{\bf 51B} (1974) 461.
\bibitem{nonRenrmlztnThrm} N.~Seiberg, {\sl Phys.~Lett.}~{\bf B318} 
(1993) 469, hep-ph/9309335.
\bibitem{BecBecSt} K.~Becker, M.~Becker and A.~Strominger, 
{\sl Nucl.~Phys.}~{\bf B456} (1995) 130. \\
J.~A.~Harvey and G.~Moore, {\it ``Superpotentials and Membrane Instantons'',\/}
hep-th/9907026.
\bibitem{hrchyBCS_NJL} J.~Bardeen, L.~N.~Cooper and J.~R.~Shrieffer,
{\sl Phys.~Rev.}~{\bf 108} (1957) 1175. \\
Y.~Nambu and G.~Jona-Lasinio, {\sl Phys.~Rev}.~{\bf 122} (1961) 345.
\bibitem{Pakvasa} S.~Pakvasa and H.~Sugawara, (to be published).
\bibitem{mtrxFrmltn} T.~Banks, W.~Fischer, S.~H.~Shenker and L.~Susskind,
{\sl Phys.~Rev.}~{\bf D55} (1997) 5112.
\bibitem{cptSpcmtrx}  N.~Seiberg, {\sl Phys.~Rev.Lett.}~{\bf 79} (1997) 3577,
hep-th/9710009.
\bibitem{mbrnThry} B.~de Witt, J.~Hoppe and H.~Nicolai, 
{\sl Nucl.~Phys.}~{\bf B305} (1988) 545.
\bibitem{3-mfdTplgy} $p=2$ case is studied in A.~J.~Casson and S.~A.~Bleiler,
{\sl Automorphism of Surface after Nielsen and Thurston,\/} Condon
Mathematical Society, Student Texts {\bf 9}, Cambridge University Press, 1988.
\bibitem{sugawaraMnbrn} H.~Sugawara, {\it ``Theory of Membrane in Heegaard 
Diagram Expansion'',\/} KEK-TH-877, \\ hep-th/0304164.
\bibitem{lensSpace} Ray-Singer torsion of $L(p,\ts q)$ is studied by C.~Nash
and D.~J.~O'Connor, {\sl J.~Math.~Phys.}~{\bf 36} (1995) 1462, hep-th/9212022.
\bibitem{SUGRA_11dim} E.~Cremmer, B.~Julia and J.~Scherk,
{\sl Phys.~Lett.}~{\bf B376} (1976) 409.
\bibitem{mirrorSymmetry} A.~Strominger, S-T.~Yau and E.~Zaslow,
{\sl Nucl.~Phys.}~{\bf B479} (1996) 243.
\bibitem{stringVac} An excellent textbook is by D.~D.~Joyce, {\sl Compact 
Manifolds with Special Holonomy,\/} Oxford Science Publications, 2000.
\bibitem{KatzVafa} S.~Katz and C.~Vafa, {\it ``Matter from Geometry'',\/}
{\sl Nucl.~Phys.}~{\bf B497} (1997) 146, hep-th/9606086.
\bibitem{AcharyaWitten} B.~Acharya and E.~Witten, {\it ``Chiral Fermions from 
Manifolds of $G_2$ Holonomy'',\/} RUNHETC-2001-27, hep-th/0109152.
\bibitem{dualForm} One of the more recent works is M.~Cveti\v c, T.~Li and 
T.~Liu, {\it ``Supersymmetric Pati-Salam Models from Intersecting D6-Branes: 
A Road to the Standard Model'',\/} hep-th/0403061.
\bibitem{nonRenTheorem} N.~Seiberg, {\it ``Exact Results on the Space of Vacua
 of Four-Dimensional SUSY Gauge Theories'',\/}
{\sl Phys.~Rev.}~{\bf D49} (1994) 6857, hep-th/9402044.
\bibitem{higherRep} For example, 
D.~G.~Lee, {\sl Phys.~Rev.}~{\bf D49} (1994) 1417.\\
X.~G.~He and S.~Meljanac, {\sl Phys.~Rev.}~{\bf D41} (1990) 1620. \\
F.~Bucclella and C.~A.~Savoy, {\sl Mod.~Phys.~Lett.}~{\bf A17} (2002) 1239,
hep-ph/0202278.
\bibitem{BBO} Vernon D.~Barger, M.~S.~Barger and P.~Ohmann,
{\sl Phys.~Rev.}~{\bf D47} (1993) 1093:~2038.
\end{thebibliography}
\end{document}